\documentstyle[12pt,aaspp4,tighten,flushrt]{article}

\def\beq{\begin{equation}}
\def\eeq{\end{equation}}
\def\ni{\noindent}
\def\.{\mathaccent 95}
\def\a{\alpha}
\def\be{\beta}
\def\bomega{{\bf \omega}}
\def\ga{\gamma}

\def\ep{\epsilon}

\def\Om{\Omega}

\def\frac#1#2{{\textstyle{{#1}\over {#2}}}}
\def\ni{\noindent}
\def\lsim{\mathrel{\rlap{\lower4pt\hbox{\hskip1pt$\sim$}}
    \raise1pt\hbox{$<$}}}
\def\gsim{\mathrel{\rlap{\lower4pt\hbox{\hskip1pt$\sim$}}
    \raise1pt\hbox{$>$}}}
\def\sqr#1#2{{\vcenter{\vbox{\hrule height.#2pt
         \hbox{\vrule width.#2pt height#1pt \kern#1pt
         \vrule width.#2pt}
         \hrule height.#2pt}}}}

\font\gkvec=cmmib10                         %for boldface lowercase
\def\bomega{\hbox{{\gkvec\char33}}}     

\def\bw{\overline \omega}
\def\bv{\overline v}
\def\ts{\times}
\def\lb{\langle}
\def\rb{\rangle}
\def\curl{\nabla {\ts}}

\def\bbv{\bar {\bf v}}
\def\bfvp{{\bf v}'}
\def\bfjp{{\bf j}'}
\def\bfv{{\bf v}}
\def\bfw{{\bomega}}
\def\bbw{\overline{\bf \bomega}}
\def\bfwp{{\bomega}'}
\def\bfbp{{\bf b}'}

\def\bbb{\bar{ \bf b}}

\def\nb{\nabla}
\def\curl{\nb\ts}
\def\div{\nb\cdot}
\def\b0{b'^{(0)}}
\def\v0{v'^{(0)}}
\def\w0{\omega'^{(0)}}
\def\bb0{\bfbp^{(0)}}
\def\bv0{\bfvp^{(0)}}
\def\bw0{\bfwp^{(0)}}
\def\bj0{\bfjp^{(0)}}

\def\bbm{\bar {\bf m}}

\def\ni{\noindent}

\begin{document}

%\draft

%\twocolumn[\hsize\textwidth\columnwidth\hsize\csname @twocolumnfalse\endcsname
\bigskip
\bigskip
\centerline{\bf Implications of mean field accretion disc theory
for vorticity and magnetic field growth} 
%\centerline{\bf and application to planet formation} 
\medskip
\centerline{Eric G. Blackman}
\centerline{Dept. of Physics \& Astronomy, University of Rochester, Rochester NY, 14627}
\centerline{and  Institute for Theoretical Physics, University of California, 
Santa Barbara, CA, 93106}
\centerline{(submitted to MNRAS)}

%\date{\today}

%\maketitle

%\vspace{3mm}
\bigskip

In addition to the scalar Shakura-Sunyaev 
$\alpha_{ss}$  turbulent viscosity 
 transport term used in simple analytic accretion disc modeling, 
a pseudoscalar  transport term also arises.
The essence of this term can be captured even in simple models
for which vertical averaging is interpreted as integration over a 
half-thickness and one separately studies each hemisphere.
The additional term  highlights 
a complementarity between mean field magnetic dynamo theory and
accretion disc theory treated as a mean field theory.
Such pseudoscalar terms have been studied, 
and can lead to  large scale magnetic field and vorticity growth.
Here it is shown that vorticity can grow
even in the simplest azimuthal and half-height integrated
disc model, for which mean quantities depend only on radius.  
The simplest vorticity growth solutions 
seem to have scales and vortex survival times  consistent 
those required for facilitating planet formation. Also it is shown that 
when the magnetic back-reaction is included to lowest order, 
the pseudoscalar driving the magnetic field growth and that driving
the vorticity growth will behave differently with respect to shearing and
non-shearing flows: the former can reverse sign in the two cases, 
while the latter will have the same sign.

%predict that vortices only grow 
%predict that when averaged over half-height and azimuthally, 
%mean b-field defined in this way grows as an $alpha^2$ dynamo,
%even if alpha-omega dynamo is operating when considering 2 dependent
%variables.
%predict that vortices should last for at least $1/\alpha_{ss}$ orbits
%when formed in this way with magnitude about $0.1\Omega$.
%These numbers are crudely consistent with 
%The location of the vortices will depend on any radial
%dependences of the $h/r$ ratio and the $q$ parameter.
%Also it is shown that while the $B$-field growth is driven by the
%difference between small scale and large scale helicity growth,
%the vorticity is driven only by the kinetic helicity. This may
%lead to the opposite sign observed for the dynamo alpha in accretion 
%discs as compared to the vorticity alpha.

%Vorticies in accretion discs are of interest for a 
%number of reasons.  Perhaps the most notable  is as a catalyst
%for planet formation in circumstellar discs.  

%Dust migration
%to the center of vortices can enable centimeter sized clumps to
%grow to meter sized clumps so that gravitational instability can
%subsequently ensue and complete the planet growth process.
%Most previous work in this context has focused on the evolution
%of pre-existing vortices, though Rossby wave vortex growth has
%been studied. 
%

%drives both the growth of vorticity and mean magnetic field, 
%an accretion disc simulation which shows evidence for a 
%mean magnetic field growth should also find vorticity growth.  
%Prvided boundary conditions are appropriates

\bigskip
\ni {\bf Key Words:} 
 
magnetic fields: MHD; instabilities; accretion
discs; turbulence; Solar system: formation; hydrodynamics

\vfill
\eject
%]

%\keywords{}

%\pacs{95.30.Q, 47.65, 52.35.R, 91.25.C}

\section{Introduction}

Analytic accretion disc theory and mean field magnetic 
dynamo theory are often thought of disjointly, but 
the two are intimately related in certain
important senses and exploring this relation has some
interesting consequences. 
In traditional $\alpha$ analytic accretion disc theory,
(e.g. Shakura \& Sunyaev 1973; Pringle 1981; Balbus-Hawley 1998)
the Reynolds+Maxwell stress tensor representing a correlation
of turbulent velocities, is replaced by $\alpha_{ss} c _s^2$, where
$c_s$ is the sound speed. In the Navier-Stokes equation, this amounts
to replacing the microphysical 
viscosity by a turbulent viscosity, $\nu =\alpha_{ss} c_s H$, 
where $H$ is the scale height.  The flow variables then being solved for 
must be interpreted as mean fields,
with gradient scales larger than those of the turbulent
motions providing the viscosity.  (Consequences of fluctuations
for variability were studied in Balbus et al. 1994; 
Blackman 1998; 2000a).
The continuity, radial momentum, angular momentum, and energy equations are
solved, often with the assumption of axisymmetry, for
radial solutions.  The magnetic field is often added
in as a pressure, rather than coupling in the magnetic induction
equations (e.g. Narayan \& Yi 1994; Narayan et al. 1998).

Indeed it is widely known that the $\alpha$ viscosity approach is
incomplete.  For more realistic discs, the dynamical
evolution of the magnetic field must be coupled to the fluid equations.
The problem is fully non-linear, and ultimately requires 
non-linear MHD turbulence simulations  understanding 
magneto-shearing instabilities etc.
(c.f. Balbus \& Hawley 1991,1998, Brandenburg et al. 1995; 
Miller \& Stone 2000).
Nevertheless, the $\alpha$ viscosity approach provides a useful
phenomenological framework for exploring solutions.  
This is  where the relationship to mean field dynamo theory enters.  

As traditional accretion disc
theory ignores the induction equation, traditional kinematic mean
field dynamo theory solves only the magnetic induction equation
(e.g. Moffatt 1978; Parker 1979; 
Krause \& R\"adler 1980; R\"adler 1999)
taking the velocity that appears in that equation as given and
not subject to the influence of the field.  Thus traditional accretion
disc theory and kinematic magnetic dynamo theory are complements
of each other. Mean field dynamo theory in its analytic regime, kinematic
or dynamic, is also subject to the complementary criticisms of accretion disc
theory in that rigorously the fully dynamical set of 
equations have to  be solved and the physics of MHD turbulence must
be understood.  For a sheared accretion disc, the problem of
accretion and the problem of field growth are one
fully coupled problem.  Disc simulations 
with appropriate  boundary
conditions may show some evidence for mean field dynamo operation 
in a turbulent accretion disc 
(Brandenburg \& Donner 1997), though higher magnetic Reynolds number
simulations are needed. Note also that the mean field 
dynamo is not to be
confused with turbulent amplification ($\equiv$ small scale dynamo).
The former models an inverse cascade of magnetic helicity
(e.g. Pouquet et al. 1976), 
not simply field line stretching on the scale of the turbulent velocities.
In addition, the mean field dynamo is often invoked to generate 
a mean magnetic flux, not just a mean field energy.

The formalism of mean field theory as applied to the induction
equation leads to both a viscosity term that  corresponds to the 
Shakura-Sunyaev $\alpha$ viscosity term, 
and an additional pseudoscalar term which can 
be similarly parameterized.  As applied to the Navier-Stokes equation,
the same formalism leads again to a Shakura-Sunyaev
$\alpha_{ss}$ type  viscosity term,
and also to a pseudoscalar helicity term.  The latter would
vanish if the vertical averaging is interpreted as integration 
over the entire disc height, but does not vanish when the integration is 
taken over a half scale height (or less) and one considers each hemisphere
independently. The pseudoscalar term has been recognized elsewhere
(e.g. Moiseev et al 1983; Frisch et al. 1987; Khomenko et al 1991;
Kitchatinov 1994ab; Tanga et al. 1996; Blackman \& Chou 1997).
Here it will be shown that it can be treated on par with the Shakura-Sunyaev
viscosity. The pseudoscalar term can lead to vorticity growth in discs.

%In the same way that the mean field magnetic dynamo characterizes
%an inverse cascade of magnetic helicity (e.g. Pouquet et al 1976),
%the psedoscalar highlights an inverse cascade of vorticity.
%Indeed 
Enstrophy exhibits an inverse cascade in 2-D turbulence
(Kraichnan \& Montgomery 1980).  The growth of vorticity in
 primarily 2-D rotating fluids has been seen in nature (e.g. Jupiter, c.f.
Ingersoll 1990; Marcus 1993) as well as in simulation
(Marcus 1990; McWilliams 1990) and experiment (e.g. 
Sommeria et al. 1988).  Statistical mechanics
approaches have been successful in modeling this (e.g.
Chavanis \& Sommeria 1998).  Vorticity growth in sheared
accretion discs is less well studied (reference added in proof: 
Godon \& Livio 2000), 
even though  
vortex evolution has been studied (e.g. Adams \& Watkins 1995;
Godon \&  Livio 1999; Chavanis 2000).
It will be later shown that an instability results in a sheared
disc even when mean quantities are integrated 
to be only a function of radius in each hemisphere.
Note here that the use of a pseudoscalar requires the turbulence to 
be 3-D even if the mean fields are taken to be 2-D.

In sections 2 and 3 the mean vorticity and mean magnetic field
equations are derived  with the back-reaction
between the two included to lowest
order, under the assumption that 1st order 
cross-correlations between fluctuating
fields and velocities are small. In section 4, 
the resulting diffusion and pseudoscalar 
transport coefficients a parameterized to be 
proportional to the sound speed and, 
growing solutions for vorticity are derived.  In section 5 	
the relation to vorticity growth for planet formation
is discussed. Why the pseudoscalar growth coefficient for vorticity can
have the opposite sign to that driving mean magnetic field growth is
also discussed. Section 6 is the conclusion.

\section{Derivation of Vorticity Equation}

%First an important comment about the magnetic field.
%While the magnetic field ultimately needs to be considered
%in more detail in the fully dynamical treatment, it is possible
%to provide some justification for ignoring it for the purposes of
%deriving the mean field vorticity equation and growth instability 
%in this work. 

%It was shown in Blackman \& Chou (1996) (c.f. eq 15 and 16 there)
%that the mean magnetic field
%couples to the mean vorticity equation only when the cross-correlations
%between magnetic and kinetic fluctuating turbulent quantities are
%finite. 
%If we ignore these correlations, then the mean field
%vorticity equation and the mean field magnetic equation essentially 
%decouple.  However the back-reaction in the transport coefficients
%is still present. This will be clearer below.

The Navier-Stokes equation with B-fields is given by  
\begin{equation}
\begin{array}{r}
\partial_{t} {\bfv}={\bfv} \ts(\nb\ts\bfv)
-{\nabla p / \rho}-{\nabla v^2/2}
 +\nu\nabla^{2}{\bf v}+ (\zeta +{1\over 3}\eta) \nabla (\div v)
+{\bf m}
+{\bf F}({\bf x}, t)+\nabla\phi,
\end{array}
\label{NS}
\end{equation}
where $\nu$ and $\zeta$ are the constant viscosity and
second viscosity and $p$ is the pressure
%$p_{\em{eff}} \equiv p+b^2/2+v^2/2$.
and ${\bf m}\equiv ({\bf B}\cdot\nabla{\bf B}-{1\over 2}\nabla {{\bf B}^2})/4\pi 
{ \rho}$.
The $\nabla\phi$ includes the potential force of gravity. 
The vorticity equation,
\begin{equation}
\partial_{t} {\bfw} =\curl(\bfv\ts\bfw)+
\nu\nabla^{2}{\bfw}
+\nabla\rho \ts \nabla p/\rho^2
+\curl{\bf F}({\bf x}, t)+\curl {\bf m}
,
\label{VORTICITY}
\end{equation}
\noindent where ${\bfw} \equiv \curl {\bfv}$,
is obtained by taking the curl of  (\ref{NS}).
Hereafter we ignore the microphysical viscosities.

Now divide quantities such as $\bfw,\ \bfv$, etc. 
into mean (indicated by an overbar or $\lb\ \rb$) and
fluctuating (indicated by prime) components
$\bfv=\bbv+\bfvp$ and  ${\bfw}=\bbw+\bfwp$ and
respectively.  The interest is in the application to  a thin disc. 
I take the average to mean full integration over azimuth,
vertical integration over the top half of the disc only,
and local radial averaging, leaving mean quantities
only a function of radius in each hemisphere, 
considering the top and bottom hemispheres of the
disc separately.  This is important because pseudoscalar averages
would vanish if the average is taken over the whole disc.
%Thus, as defined in this way,  the resulting
%means are functions only of radius.
% with $\bbv_T=\bbv+\bbv_c$.
%Here, $\bbv_c$ is a constant background velocity whose curl is
%$\bbw_c$, and $\bbv$ and $\bbw$ are the time dependent intermediate
%scale mean components whose spatial scales of variation are smaller
%spatial than that of $\bbw_c$.  
%The $\bbw_c$ reflects the background
%Keplerian disc velocity whilst the 
%$\bbw$ is the component whose growth is of interest.

The presence of ${\bf F}({\bf x},t)$ 
represents any function (e.g. forcing) not included in the other terms.
%ccretion disc would
%incorporate shear driving as well as pseudoscalar generation.  
It will drop out later. 
Note that ultimately, pseudoscalar generation  results from a vertical density
gradient combined with the shear and underlying rotation.
Later I will parameterize the 
pseudoscalar in a similar way that Shakura \& Sunyaev (1973) 
parameterize the viscosity.

%helicity to the turbulence and ensures that any growth of
%$\bbw$ represents a transfer of angular momentum from the
%turbulence.
%, consistent with angular momentum conservation. 

It is assumed that derivatives with respect to $\bf x$ or $t$
obey $\partial_{t,{\bf x}}\lb { X_i X_j} \rb =
\lb\partial_{t,{\bf x}}({X_i X_j})\rb$ and $\lb
\bar{X}_{i}X_{j}^{\prime} \rb = 0$ (Reynolds relations
(\cite{RAD80})), where $X_i={\bar X}_i+X^{\prime}_i$ are
components of vector functions of $\bf x$ and $t$.  
%%means, these hold when the 
%correlation time scales are short relative to the variation
%times of mean quantities.  
For the spatial mean,
defined by $\lb X_{i}({\bf x},t)\rb = |\zeta|^{-3} \int^{{\bf
x}+ L}_{{\bf x}-L} X_{i} ({\bf s},{\rm t}) {\rm d^3}{\bf s}$,
the relations hold when the averaging is taken over a large enough
scale, such that $l \ll |{\bf \zeta}| \ll L$, where $L\sim
{\bar v}/\nabla {\bar v}$, and $\ell \sim v'/\nabla v'$.  

Subtracting the mean of (\ref {NS}) from 
itself, and assuming $\nabla\phi'=0$, gives 
\begin{equation}
\begin{array}{l}
\partial_{t} \bfvp = 
\lb\bfvp\cdot\nabla\bfvp\rb-\bfvp\cdot\nabla\bfvp-
\bbv\cdot\nabla\bfvp
-\bfvp\cdot\nabla\bbv
 \\
-\nabla p'/\rho-\nabla (v'\cdot{\overline v})
+{\bf F}'({\bf x}, t)+{\bf m}'
%+\nu\nabla^{2}\bfvp.
\label{VPRIME}
\end{array}
\end{equation}
ignoring the viscosities.
% and assuming the barotropic equation of state
%$P=P(\rho)$.
%\noindent where $d_t\equiv \partial_t +\bbv_c\cdot\nabla$. 
%Thus the equations are cast in the frame moving with the 
%local tangential velocity.
To proceed, 
 (\ref{VORTICITY}) needs to be averaged.  To simplify,  
assume that the density has only a mean time independent 
spatially varying quantity, and ignore its fluctuating
gradients. Then, since the means are only functions of the radial
coordinate, the pressure term drops out of the vorticity. 
(The mean
pressure thus satisfies the barotropic equation of state ${\bar p}
={\bar p}({\bar \rho})$.).
This gives
\begin{equation}
\begin{array}{r}
d_t \bbw =\curl \lb\bfvp\ts\bfwp\rb
%+ \nu\nabla^{2}{\overline{\bfw}} 
+ \curl {\bf {\overline m}}
+ \bbw\cdot\nabla\bbv
%-\bbv\cdot\nabla\bbw
-\bbw\nabla\cdot\bbv, 
\end{array}
\label{OMEGABAR}
\end{equation}
\noindent where I neglected terms second order in
time-varying mean quantities and and $d_t$ indicates working in the
Local Standard of Rest (LSR) frame, i.e. that which moves with $\bar {\bf V}$. 
I have also assumed ${\bf F}={\bf F}'$.
Subtracting 
(\ref{OMEGABAR}) from  (\ref{VORTICITY}) (ignoring the pressure term
for reasons described above) gives
\begin{equation}
\begin{array}{l}
d_{t} \bfwp = \bfwp\cdot\nb\bbv-\bfwp\nabla\cdot\bbv
%-\bbv\cdot\nabla\bfwp
+\bbw\cdot\nb\bfvp-\bbw\nb\cdot\bfvp
\\
-\bfvp\cdot\nb\bbw +
\bfwp\cdot\nabla\bfvp -\bfwp\nb\cdot\bfvp 
-\bfvp\cdot\nabla\bfwp -
\curl\lb \bfvp\ts\bfwp\rb + \curl {\bf m}'
%+\nu\nabla^{2}{\bf \omega}^{\prime}
 \\
+\curl{\bf F}'({\bf x}, t).
\end{array}
\label{OMEGAPRIME}
\end{equation}
For $\bbw$ to grow, the $\nabla\times$ terms in
%(\ref{BBAR}) 
(\ref{OMEGABAR}) must be non-vanishing.  
%When the turbulence is strictly  homogeneous 
%or isotropic in $\bfvp$ these quantities vanish straight-away  
%For purely isotropic turbulence, the term
%$\lb \bfvp\times\bfwp\rb$ in  (\ref{OMEGABAR}) vanishes since it is
%the average of a vector, while $\curl \lb\bfvp\ts\bfwp\rb\propto
%\curl \nabla\lb v'^2\rb$ vanishes from 
%Reynolds rules and incompressibility.  
%so anisotropy and inhomogeneity must be
%present for nontrivial time evolution of mean fields.

Following previous work (c.f. 
Blackman \& Chou 1997; Field et al. 1999), I 
expand the turbulent quantities on the right of Eqs. 
(\ref{OMEGABAR}) to linear order in 
$\nabla \bbv$ using the equations for the fluctuating fields
%thus generalizing  the approach of 
%(\cite{FIE99}) where only the
%mean B-field was used. Unlike those previous works,  
here it is assumed that correlations of zeroth order
quantities can be functions of radius.
To find the lowest order terms,  assume 
weakly anisotropic turbulence: terms linear in the
mean shear contribute, but their averaged 0$^{th}$
order coefficients, are taken to be isotropic. 
These coefficients can be 
reflection asymmetric and radially dependent.

In order to ``ignore'' 
the terms which are products
of one zeroth order turbulent quantity with one 1st order
turbulent quantity (as I will do here) 
one assumes that these terms are small compared to 
the associated terms that involve products
of one zeroth order quantity with one mean quantity. 
For example, one would 
assume that the 1st term on the right of (\ref{OMEGAPRIME})
is greater than the sixth term on the right of (\ref{OMEGAPRIME}).
This weaker than the ``usual'' first order smoothing
approximation in that the present requirement amounts to    
${\bfwp}^{(1)}<\nabla {\overline {\bf v}}$ and  
${\bf v}^{(1)}<{\overline {\bf v}}$, rather than the usual
${\bfwp}<\nabla {\overline {\bf v}}$ or 
${\bf v}<{\overline {\bf v}}$.  The latter two conditions are  
stricter as they apply to the total fluctuating quantities rather than
only the anisotropic part.  

An alternative potential justification for dropping the troubling
terms that 
are products of one zeroth order turbulent quantity with one 1st order
turbulent quantity, is that 
they themselves likely have positive 
and negative contributions which might nearly cancel:
In a steady state, the turbulent
quantities are balanced by input and cascade.
This motivates a possible replacement of e.g. the 
sum of 6th 7th 8th and 9th 
terms by $(\chi-\xi)\bfwp^{(1)}$, where $\chi$ and $\xi$ are positive,
and their difference represents the combination of growth and decay.
(The 9th term is actually irrelevant as it vanishes when correlated
with a fluctuating quantity, which is the only context in which it will enters.)
Ignoring these terms would then amount to the assumption that 
$\zeta-\xi$ is small. Such an approach avoids 
comparing each individual ``offending'' term to those
dependent on the mean fields, since here the offending terms
would cancel themselves.

Working in the LSR frame is also important
to emphasize here.  Although the turbulence in a sheared disc is 
more than weakly anisotropic, by working in the LSR frame,
the anisotropy then manifests through terms like
the first in (\ref{OMEGAPRIME})
$\bfwp^{(0)}\cdot\nabla {\bbv}$ rather than  through terms like 
to $\bbv\cdot\nabla\bfwp^{(0)}$.
To assess the implications,  
terms like $\bfvp^{(0)}\cdot\nabla {\bbv}$ must be compared to terms like 
$\bfvp^{(0)}\cdot\nabla {\bfvp}^{(0)}$.  For accretion 
discs whose turbulence is ultimately driven by a magneto-shearing
instability, the ratio of these terms is of order 
${v^{(0)}\Omega \over ({v^{(0)}}^2/l_T)}$. But $v^{(0)}/l_T \sim \Omega$
for the magneto-shearing instability, so the anisotropy is  of order 1
in the LSR frame, compared to  $>>1$ in the lab frame.
Indeed, even in the LSR, this still means that the anisotropy should be
considered to more than linear order, but because it
is ``only'' of order 1
we expect qualitative similarities of the results to the
expansion to all orders. Blackman (2000b) gave a restricted 
approach to treating this shear anisotropy to all orders.

Using
the formal solutions for the turbulent field
%$\bfbp(t)=\bfbp(t=0)+\int d_{t'}\bfbp dt'$ 
$\bfwp(t)=\bfwp(t=0)+\int d_{t'}\bfwp dt'$, and using times
appropriately chosen such that the correlation
$\lb\bfvp(t)\ts\bfwp(0)\rb\simeq 0$, we get
%to first order in mean quantities,
\begin{equation}
\begin{array}{r}
\lb\bfvp\ts\bfwp\rb^{(1)}
=\lb\bfvp^{(0)}(t)\ts\int^t_0d_{t'} 
\bfwp^{(1)} dt' \rb +  
\lb\int_0^t d_{t'} 
\bfvp^{(1)} dt' \ts \bfwp^{(0)}(t)\rb,
\end{array}
\label{VPWP}
\end{equation}
%\begin{equation}
%\lb\bfvp\ts\bfwp\rb^{(1)}
%=\lb\bfvp^{(0)}(t)\ts\int^t_0d_{t'} 
%\bfwp^{(1)} dt' \rb
%+\lb\int_0^t d_{t'} 
%\bfvp^{(1)} dt' \ts \bfwp^{(0)}(t)\rb,
%\label{VPWP}
%\end{equation}
%\noindent with similar expressions for
%$\lb\bfbp\cdot\nabla\bfbp\rb^{(1)}$ and $\lb
%\bfvp\times\bfbp\rb^{(1)}$. 
The calculation of these  averages
requires Eqs. (\ref{VPRIME}) and (\ref{OMEGAPRIME})
%(\ref{BPRIME}) 
for the time integrands, invoking the approximations discussed above.
Using 
(\ref{VPRIME}) also requires an expression for the pressure,
which arises in (\ref{VPWP}) 
via the term
\begin{equation}
{\overline \rho}^{-1}\lb \bfwp^{(0)} (t) \ts\int_0^t \nabla p'^{(1)} dt' 
\rb,
%\,\,\, \mbox{and} \,\,\,
%\lb\bfbp^{(0)}(t)\ts\int_0^t \nabla p'^{(1)} dt'\rb.
\label{PTERMS}
\end{equation}
where the ${\overline\rho}$ is pulled out under the assumption
that it is time independent.
Using isotropy, homogeneity, and Reynolds rules, Blackman and Chou (1997)
showed that  terms of the form (\ref{PTERMS}) vanish in the derivation of the mean field equations to the first order considered.  
However in the present case, the averages of statistical
correlations can be functions of radius. Thus  
 several terms dropped from Blackman \& Chou (1997) must be considered.
Following that approach,  the energy equation can be used to obtain
\begin{eqnarray}
\nabla p'^{(1)} = \nabla p'^{(1)}(0)-\nabla\!
\int (\bfvp^{(0)}\!\cdot\!\nabla {\bar
p}+\bbv\!\cdot\!\nabla
p'^{(0)} \quad  \\ \nonumber 
+\ga p'^{(0)}\nabla\cdot\bbv+\ga
\bar{p}\nabla\cdot\bfvp^{(0)}) dt',  
\end{eqnarray}
where $\ga$ is the adiabatic index and we must now
relax the $\nabla\cdot \bfvp'^{(0)}$ constraint.
To simplify the analysis, 
I have ignored any radiative contributions to higher than zeroth order 
quantities.
%\begin{equation}
%\nabla p'^{(1)} = \nabla p'^{(1)}(t=0) - \nabla
%\int (\bfvp^{(0)}\cdot\nabla {\bar p}+\bbv\cdot\nabla
%p'^{(0)}+\ga p'^{(0)}\nabla\cdot\bbv+\ga \bar{p}\nabla\cdot\bfvp^{(0)}.
%+\ga p'^{(0)}\nabla\cdot\bfvp) dt'.  
%\end{equation}

\noindent The pressure dependent contribution to $\lb \bfvp\ts
\bfwp\rb^{(1)}$ can then be written
\begin{equation}
\begin{array}{l}
\displaystyle \lb \bw0\ts \int_{0}^{t}d_t'\nabla p' dt' \rb_{k}
 =\ep_{ijk}\ep_{ims} 
%\times
%\quad \quad \quad 
%\\[13pt] 
\displaystyle\displaystyle \int_{0}^{t}\!dt'\int_{0}^{t'}\!dt'' 
\lb (\partial_{j} {\bar v}_l \partial_l p'^{(0)}+ 
{\bar v}_l \partial_j \partial_l p'^{(0)}+   
\displaystyle  \partial_j v'^{(0)}_l 
\partial_l{\bar p}+v'^{(0)}_l\partial_j\partial_l{\bar p}
\\[13pt] 
+ \ga p'^{(0)}\partial_j\partial_l\bar{v}_l
%\\[13pt]
+\displaystyle \ga\partial_j p'^{(0)}\partial_l\bar{v}_l+
\ga{\bar p}\partial_j\partial_l v'^{(0)}_l+
\ga\partial_j{\bar p}\partial_l v'^{(0)}_l) 
%\\[13pt]
%\quad 
%\hspace{3cm}
(\partial_m v'^{(0)}_s-
\partial_s v'^{(0)}_m)\rb 
%\quad  
\displaystyle 
\\[13pt] 
={2\tau_c\over 3}\int_{0}^{t'}\!dt'
[\partial_k{\overline p}
\lb(\nabla\bfvp)^2\rb^{(0)}-\partial_k{\overline p}\lb\partial_jv'_l
\partial_lv'_j\rb^{(0)}-\bbw_k\lb\nabla p'^{(0)}\cdot\curl\bfvp^{(0)}\rb],
%{1 \over 6}\int_{0}^{t}dt' \int_{0}^{t'} dt''
%\lb \bw0(t')\cdot\bw0(t'') \rb \partial_{k}\bar{p}(t),
\end{array}
\label{P3}
\end{equation}
%\begin{equation}
%\begin{array}{ll}
%\displaystyle \lb \bw0\ts \int_{0}^{t}d_t'\nabla p' dt' \rb_{k}
%& = \displaystyle \ep_{ijk}\ep_{ims}\int_{0}^{t}dt' 
%\int_{0}^{t'}dt'' 
%\lb (\partial_{j} {\bar v}_l \partial_l p'^{(0)}+ 
%{\bar v}_l \partial_j \partial_l p'^{(0)}+ \partial_j v'^{(0)}_l 
%\partial_l{\bar p}+v'^{(0)}_l\partial_j\partial_l{\bar p} \\
%\displaystyle \: & \displaystyle\hspace{-2cm} + 
%\ga p'^{(0)}\partial_j\partial_l\bar{v}_l+
%\ga\partial_j p'^{(0)}\partial_l\bar{v}_l+
%\ga{\bar p}\partial_j\partial_l v'^{(0)}_l+
%\ga\partial_j{\bar p}\partial_l v'^{(0)}_l)
%(\partial_m v'^{(0)}_s-\partial_s v'^{(0)}_m)\rb \\
%\displaystyle \: & \displaystyle =  
%{1 \over 6}\int_{0}^{t}dt' \int_{0}^{t'} dt''
%\lb \bw0(t')\cdot\bw0(t'') \rb \partial_{k}\bar{p}(t),
%\end{array}
%\label{P3}
%\end{equation}
where it is  assumed that mean fields vary on time
scales longer than the fluctuating fields, 
and $\nabla p'^{(1)}(t=0)$ is uncorrelated with 
%$\bfbp(t)$ 
or ${\bfw}'(t)$. 
%The surviving term arises from the third term on the right hand
%side of  (\ref{P3}) and 
The vanishing of  terms to get to the last equality
follows from careful application of isotropy ({\it i.e.} rank 2 and
rank 3 averaged tensors of fluctuating 0$^{th}$ order quantities are
proportional to $\delta_{ij}$ and $\ep_{ijk}$ respectively), but 
homogeneity of the 0$^{th}$ order turbulence has not been used 
 ({\it i.e.} $\partial_{i}\langle X_{j}X_{k}\rangle^{(0)}
= 0$) nor has 
$\nabla\cdot\bfvp=0$ been used. Had homogeneity been used, then the
first term on the right of the last equality would not
have survived. 
Note however, that when put inside the curl of (\ref{OMEGABAR}),
the first two terms on the right of (\ref{P3}) vanish.
This is because the curl of the pressure gradient vanishes,
and all gradients are in the radial direction, so 
the cross product of gradients vanishes.
%when
%put inside the curl in  (\ref{OMEGABAR}).  A similar analysis
%holds for the second term in (\ref{PTERMS}) when put into
%(\ref{BBAR}); thus, the pressure does not contribute to
%(\ref{OMEGABAR}) or (\ref{BBAR}).  Inclusion of the pressure
%then simply results in an extra contribution to the coefficient
%of $\curl {\bbw}$ in the mean vorticity equation.
Finally, the last term in (\ref{P3}) vanishes from use of Reynolds
rules and isotropy, which is seen from using the chain rule
with the gradient on $p'^{(0)}$.  The pressure does not
seem to contribute to lowest order under the approximations
used herein.

Collecting all of the above,  crudely approximating time integrals 
%$\lb\bfwp\ts\bfbp\rb^{(1)},\, \lb\bfbp\cdot\nabla
%\bfbp\rb^{(1)}$ and $\lb\bfvp\times\bfbp\rb^{(1)}$ 
by factors of the correlation time $\tau_{c}$ (Ruzmaikin et al. 1988), and freely
employing Reynolds rules and incompressibility of fluctuating 
components, 
\begin{equation}
\begin{array}{l}
\displaystyle 
\curl\lb\bfvp\ts\bfwp\rb^{(1)}
= \displaystyle 
%\lb v^{0}_{i}w^{0}_{i}\rb
\curl\alpha_0\bbw 
%\lb\v0_{i}\v0_{i}\rb
-\nabla\ts(\beta_0\nabla\ts{\bbw}) 
%2\lb\w0_{i}\b0_{i}\rb\nabla^{2}\bbb  + \\
%\lb\epsilon_{ijk}\w0_{k}\partial_{i}
%\b0_{j}\rb(\curl\bbb)
%-\lb\v0_{i}\b0_{i}\rb\nabla^{2}(\curl\bbb)
\label{MESS1}
\end{array}
\end{equation}
%\noindent with similar expressions for
%$\curl\lb\bfbp\cdot\nabla\bfbp\rb^{(1)}$ and $\curl\lb
%\bfvp\times\bfbp\rb^{(1)}$
where it is assumed
${\bf F}={\bf F}^{(0)}$, and the microphysical viscosity is ignored.
The coefficients are
\begin{equation}
\begin{array}{l}
\alpha_{0} = (\tau_{c}/3)(\langle\bw0\cdot\bv0 \rb
%-2\lb\nabla p'\cdot\curl\bfvp\rb^{(0)}
 \\
%\alpha_{1} = (\tau_{c}/3)\langle\bw0\cdot\curl\bb0\rangle \\
%\alpha_{2} = (2\tau_{c}/3)\langle\bb0\cdot\bv0\rangle \\
%\alpha_{3} =(\tau_{c}/3)\langle\bb0\cdot\curl\bb0\rangle-\alpha_{0} \\
%\alpha_{3} =(2\tau_{c}/3)\langle\bb0\cdot\curl\bb0\rangle-\alpha_{0} \\
\beta_{0} =(\tau_{c}/3)\langle\bv0\cdot\bv0+\bb0\cdot\bb0\rangle \\
%\beta_{1} = (2\tau_{c}/3)\lb\bw0\cdot\bb0\rb \\
%\beta_{2} =(\tau_{c}/3)\langle\bv0\cdot\bv0+\bb0\cdot\bb0\rangle,
%\beta_{2} =(\tau_{c}/3)\langle\bv0\cdot\bv0+2\bb0\cdot\bb0\rangle,
\end{array}
\label{COEF}
\end{equation}
%Note that $\bbw_c$ and $\bbv_c$ contribute to the
%0$^{th}$ order quantities, and thus do not show up explicitly in
%(13).
where  $\bb0\equiv {\bf B}'^{(0)}/4\pi{\bar \rho}$.  
Upon substituting these into
(\ref{OMEGABAR}), 
%the curls can be pulled onto
%the  $\bbw$ and $\bbb$ from homogeneity of 0$^{th}$ order
%averages.  
\begin{equation}
\begin{array}{r}
d_{t}\bbw = \curl \alpha_{0} \bbw 
%\alpha_{1}(\curl\bbb) 
-\nabla\ts(\beta_0\nabla\ts\bbw)
%+\beta_{0}\nabla^{2}\bbw_c
%+\beta_{1}\nabla^{2}\bbb \\
%- \alpha_{2}\nabla^{2}(\curl\bbb)
+\bbw\cdot\nabla\bbv-\bbw\nabla\cdot\bbv.
%+\bbw\cdot\nabla\bbv_c+\bbw_c\cdot\nabla\bbv_c,
\end{array}
\label{WEQM}
\end{equation}
This equation presumes that 1st order cross correlation
terms vanish, that is
\beq
\lb \bfbp^{(0)}\cdot\bfvp^{(0)}\rb
=\lb{\bfwp}^{(0)}\cdot{\bf b}^{'(0)}\rb=
\lb {\bfwp}^{(0)}\cdot\curl \bfbp^{(0)}\rb=0.
\label{cond}
\eeq
When this assumption is not made, then the mean
magnetic field and mean vorticity field equations
are coupled (Blackman \& Chou 1997).
Note that since the averaging is such that the mean
fields are only functions of radius,
the penultimate term in (\ref{WEQM}) vanishes.
The last term vanishes by using the continuity equation
for a steady mean density (which has only a mean component) 
which gives 
${\overline \rho}\nabla\cdot{\overline {\bf v}}=-{\overline 
{\bf v}}\cdot\nabla{\overline \rho}$,
and then recalling that we are working in the LSR frame.
Thus 
\begin{equation}
\begin{array}{r}
d_{t}\bbw = \curl \alpha_{0} \bbw 
%\alpha_{1}(\curl\bbb) 
-\nabla\ts(\beta_0\nabla\ts\bbw)
%+\beta_{0}\nabla^{2}\bbw_c
%+\beta_{1}\nabla^{2}\bbb \\
%- \alpha_{2}\nabla^{2}(\curl\bbb)
%+\bbw\cdot\nabla\bbv_c+\bbw_c\cdot\nabla\bbv_c,
\end{array}
\label{WEQM2}
\end{equation}
is the equation to be solved.
Before proceeding to do so, I derive the magnetic field
growth equation for comparison.
%\noindent and
%\begin{equation}
%d_{t}\bbb = \alpha_{2}(\curl\bbw)+
%\alpha_{3}(\curl\bbb)+\beta_{2}\nabla^{2}\bbb+
%\bbb\cdot\nabla\bbv_c
%\label{BEQM}
%\end{equation}

\section{\bf Derivation of Magnetic Field Equation}

As mentioned above 
when cross correlations between velocity and magnetic field
components are ignored, the mean field equations for the magnetic
field and the vorticity field decouple for the case in which the
gradient of the fluctuating components of the density are ignored.
The induction equation for ${\bf B}$ is then
\begin{equation}
\partial_{t}{\bf B} =\curl(\bfv \ts{\bf B}) + 
\nu_M\nabla^{2}{\bf B},
\label{INDUCTION}
\end{equation}
\noindent Similarly, the equation for the mean 
B-field, derived by
averaging  (\ref{INDUCTION}) is
\begin{equation}
d_{t}\overline{\bf B} =\curl\lb\bfvp\ts{\bf B}'\rb
+{\overline{\bf B}}
\cdot\nabla\bbv,
%+\nu_M\nabla^{2}\overline{\bf B}
\label{BBAR}
\end{equation}
where the resistivity has been ignored and $\nabla \cdot {\overline {\bf v}}$
has been assumed.
\noindent Subtracting  (\ref{BBAR}) from (\ref{INDUCTION})
yields the equation for the fluctuating B-field 
\begin{equation}
\begin{array}{r}
d_{t} {\bf B}'= {\bf B}' 
\cdot\nb\bbv-\bbv\cdot\nabla{\bf B}' 
+{\overline {\bf B}}\cdot\nb\bfvp-{\overline {\bf B}}\nb\cdot\bfvp
-\bfvp\cdot\nb{\overline{\bf B}}
+ \\
{\bf B}' 
\cdot\nabla\bfvp -{{\bf B}'}\nb\cdot\bfvp
-\bfvp\cdot\nabla{\bf B}' 
-\curl\lb\bfvp\ts{\bf B}'\rb 
%+ \nu_M\nabla^{2}{\bf B}',
\end{array}
\label{BPRIME}
\end{equation}
again ignoring the resistivity, 
and assuming $\nabla\cdot{\overline {\bf v}}=0$.
Following the same procedure to first order in
$\overline{\bf B}$ and $\overline {\bf v}$ 
for $\lb\bfvp\ts{\bf B}'\rb$
in (\ref{BBAR}) that was followed above for 
$\lb\bfwp\ts\bfvp\rb$ 
assuming (\ref{cond}) holds, 
the mean field induction equation becomes
\begin{equation}
\begin{array}{r}
d_{t}{\overline {\bf B}}= \curl \alpha_{m0} {\overline {\bf B}} 
%\alpha_{1}(\curl\bbb) 
-\nabla\ts(\beta_{m0}\nabla\ts{\overline {\bf B}})
%+\beta_{0}\nabla^{2}\bbw_c
%+\beta_{1}\nabla^{2}{\overline {\bf B}} \\
%- \alpha_{2}\nabla^{2}(\curl\bbb)
+{\overline {\bf B}}\cdot\nabla\bbv,
%+\bbw\cdot\nabla\bbv_c+\bbw_c\cdot\nabla\bbv_c,
\end{array}
\label{BEQM}
\end{equation}
where the coefficients are
\beq
\begin{array}{r}
\alpha_{m0}=(\tau_c/3)(2\langle{\bf B}'^{(0)}\cdot
\curl{\bf B}'^{(0)}\rangle/4\pi{\bar \rho}
-\langle\bfvp^{(0)}\cdot
\curl\bfvp^{(0)}\rangle)\\
\beta_{m0}=(\tau_c/3)(2\langle{\bf B}'^{(0)}\cdot
{\bf B}'^{(0)}\rangle/4\pi{\bar \rho}+\langle\bfvp^{(0)}\cdot
\bfvp^{(0)}\rangle).
\end{array}
\label{coef2}
\eeq
and we have used $\rho=\overline \rho$. 
%Note that I do not drop the last term in (\ref{BEQM})
%as was done in for the analogous term in the vorticity equation

\section{\bf Instability growth}

%\beq%
%\partial_t{\bbw}=\curl\alpha_0\bbw
%-\curl\beta_0\curl\bbw+\bbw\cdot\nabla\bbv-\bbv\cdot\nabla\bbw
%\label{itp1}
%\eeq

Since the mean quantities are taken to be 
only a function of radius in each hemisphere, 
it is convenient to break up the vorticity equation into
toroidal and poloidal components.  Recalling that we
are working in the LSR frame so that we can ignore terms that include
factors of $\bbv$ without derivatives, the two resulting
equations that we need to solve are then of the form
\beq
d_t\bbw_\phi=\alpha_0\curl\bbw_P+\nabla\alpha_0\ts\bbw_P+\beta_0\nabla^2\bbw_\phi-\nabla\beta_0\ts(\curl\bbw_\phi)
%+\bbw\cdot\nabla\bbv_\phi
%-\bbv\cdot\nabla\bbw_\phi
\label{itp2}
\eeq
and
\beq
\partial_t\bbv_\phi=\alpha_0\bbw_\phi+\beta_0\nabla^2\bbv_\phi,
%+\bbv_P\ts\bbw_P,
\label{itp3}
\eeq
where the subscript $\phi$ indicates the toroidal component
and $P$ the poloidal component.
%Again appealing to quantities being only functions of 
%radius, the last term in (\ref{itp2}) 
%drops out: it would require a radial component of the mean
%vorticity which would in turn require non-radial derivatives of the mean
%velocity.  Thus (\ref{itp2}) 
%becomes
%\beq
%\partial_t\bbw_\phi=\alpha_0\curl\bbw_P+\nabla\alpha_0\ts\bbw_p+\beta_0\nabla^2\bbw_\phi-\nabla\beta_0\ts(\curl\bbw_\phi).
%\label{itp4}
%\eeq
%\beq
%\partial_t\bbv_\phi=\alpha\bbw_\phi+\beta\nabla^2\bbv_\phi
%\label{itp5}
%\eeq

Writing these equations for their components 
in cylindrical coordinates, and recalling that only the radial
spatial derivatives of mean quantities contribute, gives
%\beq
%\partial_t{\overline \omega}_T=-\alpha_0\nabla^2{\overline v}_T
%+{\alpha\over r^2}{\overline v}_T+\beta\nabla^2{\overline \omega}_T-{{\bar{\bf
%{\overline\omega}_{\phi}\beta\over r^2{\overline \omega}_T
%-{1\over r}\nabla\alpha\cdot\nabla(r{\overline v}_T)+{1\over r}\nabla\beta
%\cdot\nabla(r{\overline v}_T)
%\label{itp6}
%\eeq
%and
%\beq
%\partial_t{\overline v}_T=\alpha{\overline \omega}_T+\beta\nabla^2v_T-{\beta\over r^2}{\overline v}_T.
%\label{itp7}
%\eq
\beq
\begin{array}{r}
\partial_t{\overline \omega}_\phi=-\alpha_0\partial_r^2{\overline v}_\phi-{\alpha_0 \over r}
\partial_r {\overline v}_\phi+{\alpha_0\over r^2}{\overline v}_\phi+\beta_0\partial_r^2{\overline \omega}_\phi+{\beta_0\over r}\partial_r
{\overline \omega}_\phi \\
-{{\overline\omega}_{\phi}\beta_0\over r^2}-{{\overline v}_\phi\over r}\partial_r\alpha_0-(\partial_r\alpha)(\partial_r{\overline v}_\phi)+
{{\overline \omega}_\phi\over r}(\partial_r\beta_0)+
(\partial_r\beta_0)(\partial_r{\overline \omega}_\phi)
\end{array}
\label{itp8}
\eeq
and
\beq
\partial_t{\overline v}_\phi=
\alpha_0{\overline \omega}_\phi+\beta_0\partial^2_r{\overline v}_\phi+{\beta_0\over r}\partial_r{\overline v}_\phi-{\beta_0{\overline v}_\phi\over r^2}.
\label{itp9}
\eeq
We look for solutions to (\ref{itp8}) and (\ref{itp9}) of the form
\beq
{\overline w}
_\phi={\overline w}_{\phi0}(r)e^{ik_r r+nt}\ {\rm and}\   
{\overline v}_\phi={\overline v}_{\phi0}(r)e^{ik_r r+nt},
\label{solnform}
\eeq
where $k_r$, is the radial wave number, 
${\overline v}_k$ is the Keplerian speed, 
${\overline v}_{\phi0}(r)={\overline v}_k(r)\propto r^{-1/2}$,
and
${\overline w}_{\phi0}(r)\propto r^{-p}$.
%{\overline w}_{\phi0}(r_{in})
%\left({r_{in}\over r}\right)^{p}$  a
We can take $p=3/2$ since all velocities will scale with the
Keplerian speed, and the vorticity scales with the associated curl.
%However the results do not depend sensitively on $p>1$.

Since the correlation coefficients can also depend on $r$,
this dependence must be addressed.  From (\ref{itp4}) 
it is evident that $\beta_0$ is the turbulent viscosity 
which can be parameterized in the Shakura-Sunyaev (Shakura \& Sunyaev 1973) form, that is 
\beq
\beta_0=\alpha_{ss}c_s h, 
\label{ss}
\eeq
where $c_s$
is the sound speed,
$h$ is the disc half thickness, and $\alpha_{ss}$ is the 
Shakura-Sunyaev parameter.  
Similarly, $\alpha_0$ can be parameterized as 
\beq
\alpha_0=q\alpha_{ss}c_s=q\beta_0/h, 
\label{qpar}
\eeq 
where  $q$ satisfies 
$-1<q<0$ in the top half of the disc, and
$0<q<1$ in the bottom half.
Since $c_s\simeq h v_k/r$, for $h \propto r$, 
$\beta_0\propto r^{-1/2}$ and $\alpha_0 \propto r^{-3/2}$.
Taking into account the radial dependences, and plugging (\ref{solnform})
into (\ref{itp8}) and (\ref{itp9}) gives 
\beq
\left[n+\beta_0 k_r^2-{3\over 2}{\beta_0 \over r^2}+{5\over 2}
{ik_r\beta_0\over r}\right]{\overline \omega}_\phi
-\left[{3ik_r\alpha_0 \over 2r}+{3\over 2}{\alpha_0\over r^2}
+\alpha_0 k_r^2\right]{\overline v}_\phi=0
\label{itp10}
\eeq
and 
\beq
-\alpha_0{\overline \omega}_\phi+\left[n+\beta k_r^2+{3\beta_0\over 4r^2}
%-{ik\beta \over r}
\right]{\overline v}_\phi=0.
\label{itp11}
\eeq
%Since any interesting growing mode must fit inside
%the disc, let us ignore 
%factors of $1/r$ as compared to analogous terms with $k_r$.
%The equations then become
%\beq
%\left[n+\beta_0 k_r^2-{ik_r\beta_0\over r}\right]{\overline \omega}_\phi+
%\left[{ik_r\alpha_0 \over 2r}-\alpha_0 k_r^2\right]{\overline v}_\phi=0
%\label{itp12}
%\eeq
%and
%\beq
%-\alpha_0{\overline \omega}_\phi
%+\left[n+{\beta_0\over 4r^2}+\beta_0 k^2
%-{ik\beta_0 \over r}\right]{\overline v}_\phi=0.
%\label{itp13}
%\eeq
Solving for $n$ we have
\beq
n=-\beta_0 k_r^2 -{5ik_r\beta_0 \over 4r}\pm \left[
\alpha_0^2k_r^2 -{25k_r^2\beta_0^2 \over 16r^2}
+{3ik_r\alpha_0^2\over 2r}\right]^{1/2}.
\label{itp14}
\eeq
To find the real part of $n$,  first write the bracketed
term on the right of (\ref{itp14}) as 
\beq
(a+bi)=(c+di)^{1/2},
\label{itp15}
\eeq
where 
\beq
c=\alpha_0^2k_r^2 -{25k_r^2\beta_0^2 \over 16r^2}
=
\left({k_r\beta_0\over h}\right)^2\left(q^2 -{25h^2\over 16r^2}\right),
\label{aval}
\eeq
and 
\beq
d=3k_r\alpha_0^2/2r={3k_rq^2\beta_0^2\over 2rh^2}, 
\label{bval}
\eeq
and where (\ref{ss}) and (\ref{qpar}) have been used.
Note that $a,b,c,d$
(\ref{aval}), (\ref{bval})
are all real. Solving for $a$ and $b$ we get   
\beq
a=\pm {1\over \sqrt 2}[c + (c^2+d^2)^{1/2}]^{1/2}
%\simeq \pm {1\over \sqrt 2}[-a \pm a]^{1/2}.
\label{itp16}
\eeq
and
\beq
b=\pm{1\over {\sqrt 2}}[-c + (c^2+d^2)^{1/2}]^{1/2}.
\label{itp16d}
\eeq
%where the similarity in (\ref{itp16}) follows from ignoring $b^2$, 
%since it is down from the corresponding $\alpha^2$ term from $a^$ by a factor
%of $(kr)^{-2}$.
%Let us investigate separate cases when the positive vs. negative
%solution inside the brackets of (\ref{itp16d})
%that is $[-a - (a^2+b^2)^{1/2}]^{1/2}}$
%vs $[-a + (a^2+b^2)^{1/2}]^{1/2}}$. 
%For the former case, 
%If this quantity is imaginary, 
%then $h^2/r^2>q^2$ 
%then
%Taking the positive sign in (\ref{itp16}) and 
Combining with  (\ref{itp14}) we get for the real part of $n$: 
\beq
Re(n)\simeq 
%-\beta k^2 + c=
-\beta_0 k_r^2 \pm  
{1\over \sqrt 2}
\left({k\beta_0\over h}\right)
\left[q^2-{25h^2\over 16r^2} 
+ \left(q^4+\left({25 h^2\over 16 r^2}\right)^2\right)^{1/2}\right]^{1/2},
%{b\over 2\sqrt{-a}}
%=-\beta k^2 +{\beta q^2h \over 2r(h^2/r^2-q^2)}
\label{sai}
\eeq
where in calculating $(c^2+d^2)$, I have neglected
terms of order $q^2k_r^2/h^2r^2$ and $q^4/h^4 r^2$ compared
to a term of the order $q^4k_r^2/h^4$, on the assumption that the 
relevant growth modes have 
$k >> 1/r$ (verified later).

For $q < {5h/4r}$, only the negative term in (\ref{sai}) survives
and we have only decaying solutions.
For $q >> {5h/4r}$ we have 
\beq
Re(n)=\pm k_r\beta_0q/h -k_r^2\beta_0.
\label{lim1}
\eeq
Taking the postive sign, growth occurs for $k_r < q/h$ and 
the maximum growth rate
occurs at $k_{max}=q/2h$ with maximum growth rate
$q^2\beta_0/4h^2$.
The growth is strongly dependent on $q$.

%Since $4r^2k^2>>1$, 
%Since growth modes must fit within the radius
%of the disc, we can ignore the last term in the brackets as 
%per earlier assumptions.

A similar analysis shows that the 
imaginary part of $n$ is given by 
\beq
Im(n)\simeq -{5k_r \beta_0\over 4r} 
\pm  
{1\over \sqrt 2}
\left({k_r\beta_0\over h}\right)
\left[{25h^2\over 16r^2}-q^2 
+ \left(q^4+\left({25 h^2\over 16 r^2}\right)^2\right)^{1/2}\right]^{1/2},
\label{sai2}
\eeq
for which only the first term on the right contributes when $q>>5h/4r$.
When $q<<5h/4r$ the positive solution gives zero imagnary part,
whilst the negative soultion gives $Im(n)=-5k_r\beta_0/2r$.

%Note that the magnetic field 
%equation (\ref{BEQM2}) has the same form as 
%the equation (\ref{WEQM2}) which we just solved. 
%Similar solutions can be obtained for the field (the $\alpha^2$ dynamo)
%with $q$ replaced by $q_m$ and
%$\alpha_{0}$ and $\beta_{0}$ replaced by 
%$\alpha_{m0}$ and $\beta_{m0}$.  An important consequence of
%the difference between $\alpha_{m0}$ and $\alpha_{0}$ will
%be discussed in section 5.2.

\section{\bf Discussion}

\subsection{\bf Planet formation}

The simple approach in the previous section turns out to
provide vorticity growth with numbers comparable 
to what is required by planet formation.

Studies of planet formation have invoked the idea
of trapping dust particles in vortices, to provide the required
 agglomeration of material (Barge \& Sommeria 1995; Tanga et al 1996; 
Hodgson \& Brandenburg 1998
Godon \& Livio 1999; Bracco et al. 1999 
Chavanis 2000)  
In all of the above references,  anti-cyclonic vortices (vortices opposed to
the underlying rotation) are the only ones of interest for planet formation.
Consider a vortex centered in the disc at radius $r_0$.
Dust particles entering the vortex from $r>r_0$
lose angular momentum and fall inward toward the vortex core, 
while those entering from $r<r_0$  gain  angular
momentum and move outward, also  toward the radius of the vortex core.
While there is a limited size range of dust particles that
couple to the vortex, the 
result is migration of material to the center for suitable
anti-cyclonic vortices.  

To see that the discussion of the previous section leads
to anti-cyclonic vortex growth,  
first note that ultimately,  
$\alpha_0$ results from a vertical density gradient and the Coriolis force
(hidden in the averaging rather than explcitly considered), so 
$\alpha_0$  is positive in the northern hemisphere
and negative in the southern hemisphere. 
Now the dominant growth term  
on the right  of (\ref{itp8}) is the first term.  That term is thus 
negative in the north and positive in the south.  Thus
the toroidal vorticity grows a negative contribution in the north
and positive contribution in the south.  This means that in (\ref{itp9}),
the toroidal velocity and thus the poloidal vorticity
grow negative contributions (anti-cyclonic) in the north and also
in the south.

%Next, if we infer the value of $q$ from disc simulations 
%(Brandenburg \& Donner 1997) roughly the kinematic
%value of $\alpha_\sim c_s(l/h)^2=\alpha_{ss}$ is 
%To assess the value of the coefficients $\alpha$ and $\beta$ we
%appeal to the results of simulations for accretion discs.
%Taking the rough scaling $\nu=v_T l=\alpha c_s H=v_T^2/\Omega$, where the
%latter equality comes from the fact that the 
%that the magneto-shearing instability growth times is of order
%the largest eddy turnover time, which is of order the rotation time.
%For the pseudoscalar term we use the parameterization 
%$\alpha=l^2/H^2 \omega_c H$

If we consider the case
for which $q\sim 1 >> h/r$, then we have $k_{max}\sim 1/2h$ and
$n_{max}=\beta/4h^2$ from the previous section.  
When the vortex reaches saturation, 
its characteristic evolution time would then be given by the
oscillating part of the solution.  From  the $q>>5h/4r$ limit of 
(\ref{sai2}) 
we have
\beq
\tau_{v}\gsim 2rh/\beta_0 \sim 
{2r \over \alpha_{ss}c_s}\sim {2r\over h  \alpha_{ss}\Omega},
\label{vsurv}
\eeq
where we have used $k\sim k_{max}$.
For $\alpha_{ss}\sim 0.01$ and $h/r\sim 0.1$, this limit would give
$\gsim 300$ orbital periods. This is 
consistent with what is required 
to form planets (c.f. Godon \& Livio 1999).

The saturated energy density associated with the
mean field vortex should be no greater than that associated with the
turbulent energy density.  Using the fact that 
the growth time  
for the magneto-shearing instability driving the growth of turbulence
in accretion discs is of order the dominant eddy turnover time,
and is equal to the rotation period 
$\Omega^{-1}$, we have for the viscosity $\beta_0 =\alpha_{ss}c_s
h\sim v_T^2/\Omega$, where $v_T$ is the dominant turbulent speed.
Thus   $v_T\sim \alpha_{ss}^{1/2}c_s$ and a vortex with 
the associated velocity for a mode of $k_r\sim q/h$  
would have 
\beq
\omega \sim \alpha_{ss}^{1/2}c_s k_{max}
\simeq \alpha_{ss}^{1/2}\Omega |q|\sim 0.1\Omega,
\label{vortval}
\eeq
for $q^2>>h^2/r^2$, and $\alpha_{ss}\sim 0.01$.
This is typical of that employed by 
%
%Survival of O$(10^2)$ 
%is consistent with that employed 
by Godon \& Livio (1999).
%and that required by  Barge \& Sommeria (1995).

Rewriting Godon \&  Livio's (1999) allowed range in  
size of  surviving sub-sonic vortices we have
\beq
{\beta_0 \over v_T} \le k^{-1} \le {v_T \over  \partial_r \Omega},
\label{cond2}
\eeq
where the lower bound is a viscous length, 
%$\Omega$ is the background disc (e.g. Keplerian) angular speed,
the upper bound ensures that the vortex is 
not subject to destruction by shear, and we have used 
$v_T$ for the vortex speed.
Using the relations for $v_T$ and $\beta_0$ above,
we get the condition
$\alpha_{ss}^{1/2}h\le k^{-1} \le \alpha_{ss}^{1/4}(rh)^{1/2}$.
For $k\sim {h}^{-1}$ the lower inequality is always satisfied,
and the upper inequality is satisfied for $h/r \le \alpha^{1/2}$
which is consistent with the range allowed  at least
in FU Orionis for which $\alpha_{ss}$ could
be even as large as 0.1 (Bell et al. 1995).

These values are encouraging but 
it is also important to understand why a vortex core appears
at a particular radius.  For this simple-minded approach,
the radial dependence of the growth rate could enter
through any hidden dependences in $q$. 
One could extremize $(q-5h/4r)$ to find the radius corresponding
to the maximum. However, it is likely that the favorable radius of planet
formation ultimately has much to do with the ambient dust  
concentration and size distribution as a function of radius 
in a stellar nebula (Barge \& Sommeria 1995).

It is very important to note that the overly simplified  
approach herein does not 
locate the vortices on azimuth, as the azimuth is averaged over.
Thus for a given radius, only the net vorticity is measured.
The location of the vortex, or the number of the vortices is
not determined.

%growth rate

%scale of vortices

%saturation strength

%comparison to what is required for planet formation from tanga and
%sonneria

%decay rate, lifetime of vortices.

\subsection{\bf Implications for accretion disc dynamos}

The growth of mean vorticity and mean magnetic field require
the presence of a pseudoscalar.  It is interesting to compare
the resulting pseudoscalars $\alpha_0$ and $\alpha_m$
by including  the magnetic back-reaction to lowest order.

Note that $\alpha_0$ is proportional only to the negative of 
kinetic helicity 
whilst $\alpha_m$ represents the residual between this and a
current helicity contribution. 
(The relation between this and the form in Pouquet et al. (1976)
is discussed in Field et al. 1999).  This can in principle be tested
in simulations, though one must be careful not to use periodic 
boundary conditions over the simulation box half-thickness over
which the vertical averaging is done.
In addition, it was noted in Brandenburg \& Donner (1997)
and Brandenburg (1999) that $\alpha_m$ seems to have the opposite
sign than expected from just the kinetic helicity part. It was suggested
that the shear may play the role of flipping the sign of a
rising loop of field (see Brandenburg \& Donner 1997).
Here we can go a step further and note that it is actually
the current helicity term that will be affected by that sign:
imagine a seed toroidal field from which a loop rises
and twists less than 1/4 turn, in the direction
opposite to the underlying rotation.  This would have positive
current helicity.  But now as the Keplerian shear takes over, the loop
rotates past a 1/4 turn, and the sign of the current helicity
changes.
By contrast,  the sign of the vorticity growth coefficient $\alpha_0$ is
the same for a Keplerian and non-shearing
flow, since the kinetic helicity does not change
sign. Thus in magneto-shearing simulations in which the
turbulent magnetic
energy slightly dominates the turbulent kinetic energy, the
current helicity term may dominate $\alpha_m$.

\subsection{\bf Implications for accretion disc modeling}

The growth of vorticity as modeled herein 
may have implications for angular momentum transport
in accretion discs. Indeed the presence of anti-cylconic
vortices in the underlying flow transports angular momentum out
of the ambient flow.  Note however, that 
in the present treatment we have taken
a base turbulent state that acts as a viscosity,
so angular momentum is being transported also by the underlying turbulence.
It is not entirely clear what the consequences of the
vorticity growth would be for angular momentum transport,
above and beyond their ability to agglomerate dust particles
in the accretion flow. Such an additional effect of $\alpha_0$ needs to be
calculated, since the pseudoscalar term is as naturally present as the
viscosity term when integrating only over the half-thickness
of the disc.

Note that 
the generation of Rossby wave vortices has been considered
when there is no turbulence present and leads to 
outward transport of angular momentum independently of any magnetic field.
(Lovelace et al. 1999; Li et al 1999). 
%Here we have just considered vortex generation
%when the accretion disc is already transporting angular momentum 
%at some rate by e.g. Balbus-Hawley (1991,1998) induced turbulence.   

Additional 
consequences of vortex generation 
for high energy accretion discs may include concentrating disc 
emission into collimated  beams (\cite{ACO90,ABR92}).  
X-ray iron line modeling could include the presence of such
vortices to diagnose their presence.
Also, Yoshizawa \& Yokoi (1993) discussed how the interplay of magnetic
field and vorticity can lead to generation of collimated
large scale jets.  
%Homogeneous coupled vorticity-magnetic dynamo
%was discussed in Blackman \& Chou (1997).

%an equation similar to (\ref{BEQM}) 

\section{\bf Conclusions}

Standard axisymmetric Shakura-Sunyaev type 
turbulent accretion disc theory should always be regarded 
as mean field theory. 
%Some implications of this for interpreting 
%variability were investigated by Ba
%Blackman (1998,2000). 
When the vertical averaging
is taken to be integration over 1/2 the disc scale height,
the presence of a pseudoscalar transport coefficient in addition to the usual
scalar diffusion term should survive.
This pseudoscalar term allows vorticity growth in each
hemisphere even when mean quantities are only a function of radii there.
The simplest growth rates, survival times and spatial scales seem to be
consistent with that required by planet formation studies.
%The growth discussed herein is the simplest approach for finding
%vortex growth in an already turbulent disc, unlike the
%growth of Rossby vortices (Lovelace et al. 1999; Li et al. 1999)
%for a non-turbulent disc.
The simple pseudoscalar term driving vorticity growth differs from
that which drives mean magnetic field growth in that the latter,
unlike the former, could  have an opposite sign for cases of a rigid rotator
vs. a sheared rotator such as an accretion disc.

All of these points could be tested numerically, though one must
be very careful about boundary conditions.
Future analytic work can investigate the implications of 
including the pseudoscalar term in a generalization 
of the Shakura-Sunyaev/slim disc models.
Azimuthal dependence must also be considered.

\ni {\bf Acknowledgments}:
Thanks to J.-H. Yang for a thorough study of the original version's
calculations and for corrections. Thanks to A. Brandenburg
for comments. Thanks to the ITP and to the 
Aspen Center for Physics for facilities and interactions.  
This work was supported by NSF grant PHY94-07914 while at ITP,
and by DOE grant DE-FG02-00ER54600.

\end{document}